\documentclass[12pt]{article}
\usepackage{amsfonts,amsmath,amssymb,epsf,framed}
\usepackage{amsmath,braket}
\usepackage{graphicx,hyperref,color,comment}
\usepackage{cite}

\usepackage{subfigure}

\edef\restoreparindent{\parindent=\the\parindent\relax}
\usepackage{parskip}
\restoreparindent

\hypersetup{
    bookmarks=true,         
    unicode=false,          
    pdftoolbar=true,        
    pdfmenubar=true,        
    linktocpage=true,       
    pdffitwindow=false,     
    pdfstartview={FitH},    
    pdfnewwindow=true,      
    colorlinks=true,       
    linkcolor=blue,          
    citecolor=blue,        
    filecolor=blue,      
    urlcolor=blue           
}
\numberwithin{equation}{section}									

\newcommand{\la}{\langle}
\newcommand{\ra}{\rangle}

 \let\b=\beta  \let\d=\delta \let\e=\epsilon  \let\g=\gamma \let\h=\eta  \let\l=\lambda \let\m=\mu \let\n=\nu
 \let\p=\phi \let\r=\rho \let\s=\sigma \let\t=\tau \let\th=\theta   \let\x=\xi  
  \let\G=\Gamma \let\L=\Lambda       
\def\nn{\nonumber}
\def\inf{\infty}

\def\pa{\partial}

\begin{document}

\begin{titlepage}
\thispagestyle{empty}

\vspace*{-2cm}
\begin{flushright}
UT-Komaba/25-8\\
\vspace{2.0cm}
\end{flushright}

\bigskip

\begin{center}
\noindent{{\Large \textbf{Time-like Janus Solution\\[14pt]-- holographic global quantum quench --}}}\\
\vspace{1.5cm}

Kenta Suzuki
\vspace{1cm}\\

{\it Graduate School of Arts and Sciences, University of Tokyo\\[4pt] Komaba, Meguro-ku, Tokyo 153-8902, Japan}
\vspace{1mm}

\bigskip \bigskip
\vskip 5em
\end{center}

\begin{abstract}
We construct a time-like Janus solution, which is mediated by a time-dependent dilaton field in asymptotic AdS spacetime.
This solution breaks the null energy condition, but we argue that it is nevertheless useful as a toy model of holographic global quantum quench.
The dual CFT is given by conformal perturbation theory, where the primary scalar operator that is dual to the bulk dilaton field is coupled with a global-quench-type time-dependent source.
We compute one-point functions of the scalar operator and the stress-energy tensor, and confirm that the results are consistent with the proposed CFT picture.
We also evaluate the holographic entanglement entropy for late time after the global quench.
Finite temperature version of the solution, which can be interpreted as a time-dependent black hole, is also discussed.

\end{abstract}

\end{titlepage}

\newpage

\tableofcontents

\section{Introduction}
\label{sec:introduction}

The AdS/CFT correspondence \cite{Maldacena:1997re,Gubser:1998bc,Witten:1998qj} beautifully exemplifies the idea of holography \cite{tHooft:1993dmi,Susskind:1994vu}
in clarity and precision, more powerfully than any other examples of the gauge/gravity correspondence.
It provides an explicit and promising approach to quantum gravity by rewriting it in terms of a microscopic and non-gravitational theory.
It can also provide a novel approach for studying the strong-coupling limit of a large class of quantum field theories, thanks to its strong-weak duality nature.

The Janus solution \cite{Bak:2003jk,Freedman:2003ax,Bak:2007jm,DHoker:2007zhm} is an exact solution of the Einstein-dilaton theory,
which has a geometry of AdS${_d}$-slicing domain wall in AdS$_{d+1}$ spacetime, mediated by a massless dilaton field depending on the slicing coordinate.
The asymptotic value of the dilaton field approaches a constant in one half of the boundary space and a different value in the other half.
Hence, this solution provides one of the simplest examples of the AdS/ICFT correspondence. 
This solution is used to compute the holographic boundary/interface entropy \cite{Azeyanagi:2007qj,Chiodaroli:2010ur,Goto:2020per},
holographic one-point functions in the AdS/BCFT correspondence \cite{Suzuki:2022xwv,Izumi:2022opi},
and to construct a traversable wormhole \cite{Kawamoto:2025oko}.
With its vast application of the Janus solution, it is interesting to consider another type of Janus deformation.

The previously studied Janus solution is a space-like solution, in the sense that the dilaton field has a spatial dependence.
In this paper, we construct a time-like Janus solution whose dilaton field has a time-dependence and no spatial dependence.
A time-like Janus solution was previously considered in \cite{Kanda:2023jyi}, but their solution contains a naked singularity.
In order to avoid this naked singularity, we perform an analytical continuation of the Janus deformation parameter to a pure imaginary value.
This leads to a breaking of the null energy condition.
We nevertheless argue that our solution is useful as a toy model of holographic global quantum quench in the context of the AdS/ICFT correspondence.
A Janus deformation of de Sitter spacetime was also discussed in \cite{Bak:2024bps},
which can also be thought of as some kind of time-like Janus solution, but it is different from our time-like Janus solution.

In the rest of this section, we review two types of slicing of the empty AdS$_3$ spacetime.
In section~\ref{sec:solutions}, we construct our time-like Janus solution in asymptotically AdS$_3$ spacetime.
Then we discuss application of the time-like Janus solution for the AdS/ICFT
by computing one-point functions in section~\ref{sec:one-point} and holographic entanglement entropy in section~\ref{sec:entropy}.
In section~\ref{sec:one-point}, we also propose that the dual ICFT is given by conformal perturbation theory with a global-quench-type time-dependent source.
We confirm this proposal by computing the one-point functions and comparing with the bulk results.
The time-like Janus solution breaks the null energy condition, but nethertheless we will show that at least it is stable under a probe scalar perturbation in section~\ref{sec:stability}.
We consider a generalization to the finite temperature in section~\ref{sec:finite} and higher dimension in section~\ref{sec:general-d}.
We will give our conclusion with discussions in section~\ref{sec:conclusions}.
Appendix~\ref{app:euclidean} provides some detail explanations about Euclidean solutions.
Appendix~\ref{app:integrals} summarizes detail computations in the conformal perturbation theory used in section~\ref{sec:one-point}.
Appendix~\ref{app:extremal} discuss the extremal surface for the computation of the holographic entanglement entropy.

\begin{figure}[t!]
	\begin{center}
		\scalebox{1.0}{\includegraphics{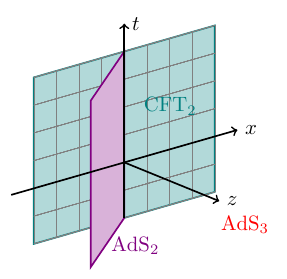}} \qquad \quad \raisebox{10pt}{\scalebox{1.0}{\includegraphics{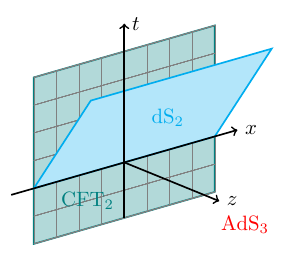}}}
	\caption{A schematic picture of AdS$_2$ slicing and dS$_2$ slicing of AdS$_3$ spacetime.}
	\label{fig:slicing}
	\end{center}
\end{figure}

\begin{figure}[t!]
	\begin{center}
		\scalebox{1.0}{\includegraphics{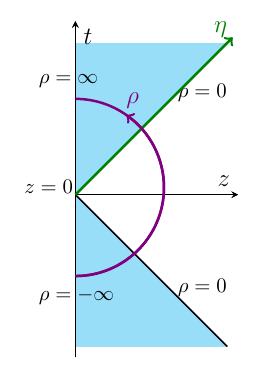}} \qquad \qquad \scalebox{1.0}{\includegraphics{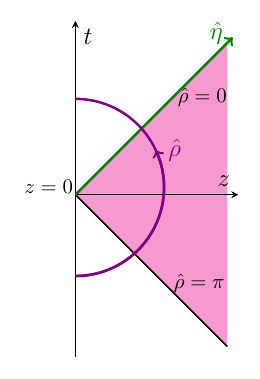}}
	\caption{The coordinates ($\r, \h$) cover the blue region in the left panel, and the coordinates ($\hat{\r}, \hat{\h}$) covers the pink region in the right panel.}
	\label{fig:coordinates}
	\end{center}
\end{figure}

Before moving on to the construction of our time-like Janus solution, let us here review the AdS$_2$ and dS$_2$ slicing of the empty AdS$_3$ spacetime.
A schematic picture of these slicing coordinates are depicted in figure~\ref{fig:slicing}.
For the Poincare coordinates of empty AdS$_3$
	\begin{align}
		ds_3^2 \, = \, \frac{-dt^2+dx^2+dz^2}{z^2} \, , 
	\label{eq:poincare}
	\end{align}
the AdS$_2$ slicing coordinates are obtained by coordinate transformations
	\begin{align}
		x \, = \, y \, \tanh \r \, , \qquad z \, = \, \frac{y}{\cosh \r} \, , 
	\end{align}
which lead to 
	\begin{align}
		ds_3^2 \, = \, d\r^2 + \cosh^2 \r \left( \frac{-dt^2 + dy^2}{y^2} \right) \, .
	\end{align}
The space-like Janus solution has this type of geometry mediated by a dilaton field $\p = \p(\r)$.
On the other hand, from the Poincare metric (\ref{eq:poincare}) the dS$_2$ slicing of AdS$_3$ spacetime for a region $t>z$ is obtained by coordinate transformations
	\begin{align}
		t \, = \, \h \, \coth \r \, , \qquad z \, = \, \frac{\h}{\sinh \r} \, , 
	\label{eq:coord-transf}
	\end{align}
where $\h>0$ and $\r>0$, which lead to 
	\begin{align}
		ds_3^2 \, = \, d\r^2 + \sinh^2 \r \left( \frac{-d\h^2 + dx^2}{\h^2} \right) \, .
	\label{eq:ds2-slicing}
	\end{align}
This metric covers the upper triangle of the left panel in figure~\ref{fig:coordinates}.
For the $t<-z$ region, we need to use coordinate transformations
	\begin{align}
		t \, = \,  \h \, \coth \r \, , \qquad z \, = \, - \, \frac{\h}{\sinh \r} \, , 
	\end{align}
with $\r<0$, which again lead to the metric (\ref{eq:ds2-slicing}), but it now covers the lower triangle of the left panel in figure~\ref{fig:coordinates}.
We would like to construct a time-like Janus solution, which has this type of geometry mediated by a dilaton field $\p = \p(\r)$.
In the following discussion, we mainly consider $t>0$ (i.e. $\r>0$) region, otherwise stated explicitly.

\section{Time-like Janus Solutions}
\label{sec:solutions}
In this section, we study the three-dimensional solution and postpone our discussion for higher dimension in section~\ref{sec:general-d}.
As in the original Janus solution \cite{Bak:2007jm}, we study the Einstein-dilaton theory in three-dimensional AdS spacetime:
	\begin{align}
		I \, = \, \frac{1}{16\pi G_N} \int d^{3}x \sqrt{-g} \Big[ R \, - \, g^{\mu\nu} \pa_\mu \p \pa_\n \p \, + \, 2 \Big] \, .
	\label{eq:I_tot}
	\end{align}
The equations of motion derived from (\ref{eq:I_tot}) are the Einstein and Klein-Gordon equations
	\begin{gather}
		R_{\m\n} \, + \, 2 g_{\m\n} \, = \, \pa_\m \p \pa_\nu \p \, , \label{eq:einstein-eq} \\
		- \frac{1}{\sqrt{-g}} \pa_\m \big( \sqrt{-g} g^{\m\n} \pa_\n \p \big) \, = \, 0 \, .
	\end{gather}

In order to search for a time-like Janus solution with the dS$_2$ slicing, we employ the following ansatz 
	\begin{gather}
		ds_3^2 \, = \, d\r^2 + f(\r) ds_{\textrm{dS}_2}^2 \, , \qquad ds_{\textrm{dS}_2}^2 \, = \, \frac{-d\h^2 +dx^2}{\h^2} \, , \label{eq:ansatz1} \\
		\p \, = \, \p(\r) \, , \label{eq:ansatz2}
	\end{gather}
with unknown functions $f(\r)$ and $\p(\r)$.
Upon this ansatz, the Einstein and Klein-Gordon equations are reduced to 
	\begin{gather}
		2ff'' \, - \, f'^2 \, - \, 4 f^2 \, + \, 2 f^2 \p'^2 \, = \, 0 \, , \\[2pt]
		f'^2 \, - \, 4 f \, - \, 4 f^2 \, - \, 2 f^2 \p'^2 \, = \, 0 \, , \label{eq:fp-eq}\\[2pt]
		\pa_\r \big(f \p' \big) \, = \, 0 \, ,
	\end{gather}
where the prime denotes a derivative with respect to $\r$.
The Klein-Gordon equation leads to 
	\begin{align}
		\p' \, = \, \frac{\hat{\g}}{f} \, ,
	\end{align}
with an integration constant $\hat{\g}$.
Provided with this result of the Klein-Gordon equation, the remaining two Einstein equations are not independent of each other, and they are also solved by 
	\begin{align}
		f(\r) \, = \, \frac{-1+\sqrt{1-2\hat{\g}^2}\cosh(2\r)}{2} \, .
	\end{align}
When $\hat{\g}=0$, this agrees with the empty AdS$_3$ as $f(\r)=\sinh^2\r$. This time-like Janus solution was previously considered in \cite{Kanda:2023jyi}.
Taking a trace of the Einstein equation (\ref{eq:einstein-eq}), the Ricci scalar is given by
	\begin{align}
		R \, = \, \big( \phi' \big)^2 - \, 6 \, = \, \left( \frac{\hat{\g}}{f} \right)^2 - \, 6 \, .
	\end{align}
From this expression, one can see that when $\hat{\g}^2>0$ the solution develops a naked singularity.
In order to avoid a naked singularity, in the following we focus on the $\hat{\g}^2<0$ case, where there is no naked singularity.
Therefore, it is more convenient to introduce $\hat{\g}=i \g$ with $\g \in \mathbb{R}$, such that
	\begin{gather}
		\p' \, = \, \frac{i \g}{f} \, , \qquad f(\r) \, = \, \frac{-1+\sqrt{1+2\g^2}\cosh(2\r)}{2} \, , \qquad (\g \in \mathbb{R}) \label{eq:f(rho)} \\[2pt]
		R \, = \, \big( \phi' \big)^2 - \, 6 \, = \, - \left( \frac{\g}{f} \right)^2 - \, 6 \, .  \qquad 
	\end{gather}
In the following discussion, we consider the $\g>0$ case. The $\g<0$ case is related by a complex conjugation.
We note that $f(\r)>0$ for any value of $\r \in \mathbb{R}$. Finally, the dilaton solution is given by
	\begin{align}
		\p(\r) \, = \, \p_0 \, + \, \frac{1}{\sqrt{2}} \log \left( \frac{1+\sqrt{1+2\g^2}-i \sqrt{2}\g \coth \r}{1+\sqrt{1+2\g^2}+i \sqrt{2}\g \coth \r} \right) \, .
	\label{eq:phi-sol}
	\end{align}
In summary, our time-like Janus solution is given by
	\begin{gather}
		ds_3^2 \, = \, d\r^2 + f(\r) ds_{\textrm{dS}_2}^2 \, , \qquad ds_{\textrm{dS}_2}^2 \, = \, \frac{-d\h^2 +dx^2}{\h^2} \, , \label{eq:Janus-metric} \\[2pt]
		f(\r) \, = \, \frac{-1+\sqrt{1+2\g^2}\cosh(2\r)}{2} \, , \qquad (\g > 0) \label{eq:fsol} \\[2pt]
		\p(\r) \, = \, \p_0 \, + \, \frac{1}{\sqrt{2}} \log \left( \frac{1+\sqrt{1+2\g^2}-i \sqrt{2}\g \coth \r}{1+\sqrt{1+2\g^2}+i \sqrt{2}\g \coth \r} \right) \, .
	\end{gather}
We note that the Wick rotation of the deformation parameter has nothing to do with a Wick rotation of spacetime coordinates.
Even after double Wick rotations from the Euclidean version of the usual space-like Janus solution, we still need a Wick rotation of the defomation parameter.
We will discuss more in detail the Euclidean solutions in Appendix~\ref{app:euclidean}.

This time-like Janus solution breaks the null-energy condition.
A null vector is given by
	\begin{align}
		N^\m \, = \, \left\{1, \, \frac{\h}{\sqrt{f}}, \, 0 \right\} \, ,
	\end{align}
and the null-energy condition is 
	\begin{align}
		R_{\m\n} N^\m N^\n \, = \, \frac{f'^2-2f-ff''}{2f^2} \, = \, \p'^2 \, = \, - \left( \frac{\g}{f} \right)^2 \, .
	\end{align}
With $\g \in \mathbb{R}$, the solution (\ref{eq:f(rho)}) gives $R_{\m\n} N^\m N^\n \le 0$, where the equality is saturated by the empty AdS$_3$ ($\g=0$).
Therefore, this time-like Janus solution is energetically unstable, and one might wonder what the usefulness of this solution is.
Here let us point out a similarity with traversable wormholes.
This construction of the solution is similar to the construction of traversable wormholes \cite{Gao:2016bin} in the sense that we modify the geometries by tuning on the negative energy.
Furthermore, the construction of wormholes by tuning imaginary sources for marginal operators was recently discussed in \cite{Garcia-Garcia:2020ttf, Kawamoto:2025oko}.
Even though this time-like Janus solution breaks the null-energy condition, we will argue that it is nevertheless useful as a toy model of holographic global quantum quench from the next section.

\subsection{Solution for the other patch}
Before discussing an application for the AdS/CFT correspondence, let us also consider the other patch.
As we mentioned, the solution presented above only covers the blue patch depicted in the left panel in figure~\ref{fig:coordinates}.
In this patch, $\h$ is a time-like coordinate and $\r$ is a space-like coordinate.
The other patch in the right panel in figure~\ref{fig:coordinates} can be thought of a doubly Wick rotation $\h$ and $\r$.
Therefore, we introduce $\r = i \hat{\r}$ and $\h = i \hat{\h}$.

From the Poincare coordinates of AdS$_3$, if we use coordinate transformation
	\begin{align}
		t \, = \, \hat{\h} \, \cot \hat{\r} \, , \qquad z \, = \, \frac{\hat{\h}}{\sin \hat{\r}} \, , 
	\end{align}
we arrive at
	\begin{align}
		ds_3^2 \, = \, - d\hat{\r}^2 + \sin^2 \hat{\r} \left( \frac{d\hat{\h}^2 + dx^2}{\hat{\h}^2} \right) \, .
	\end{align}
Therefore, now $\hat{\r}$ is a time-like coordinate and $\hat{\h}$ is a space-like coordinate.

Now we consider an ansazt
	\begin{gather}
		ds_3^2 \, = \, -d\hat{\r}^2 + f(\hat{\r}) ds_{\textrm{EAdS}_2}^2 \, , \qquad ds_{\textrm{EAdS}_2}^2 \, = \, \frac{d\hat{\h}^2 +dx^2}{\hat{\h}^2} \, ,  \\
		\p \, = \, \p(\hat{\r}) \, ,
	\end{gather}
then the solution is given by
	\begin{gather}
		f(\hat{\r}) \, = \, \frac{1-\sqrt{1-2\g^2}\cos(2\hat{\r})}{2} \, , \qquad (\g \in \mathbb{R}) \\[2pt]
		\p(\hat{\r}) \, = \, \p_0 \, + \, \frac{1}{\sqrt{2}} \log \left( \frac{1+\sqrt{1-2\g^2}-i \sqrt{2}\g \cot \hat{\r}}{1+\sqrt{1-2\g^2}+i \sqrt{2}\g \cot \hat{\r}} \right) \, .
	\end{gather}
This solution covers the pink patch in the right panel of figure~\ref{fig:coordinates}.
Nevertheless, for an application for AdS/CFT correspondence, we mostly need the previous solution (\ref{eq:Janus-metric}), since the blue patch covers almost all asymptotic AdS boundary, except $t=0$ line.

\section{One-point Functions}
\label{sec:one-point}
Now we consider the AdS/CFT correspondence with the bulk solution given by the ones discussed in the previous section.
We note that the asymptotic value of the solution (\ref{eq:phi-sol}) is given by
	\begin{align}
		\p_+ \, &= \, \p_0 \, + \, \frac{1}{2\sqrt{2}} \log \left( \frac{1-i \sqrt{2}\g}{1+i \sqrt{2}\g} \right) \, , \qquad (\textrm{for} \ \ t>0) \label{eq:phi_+} \\	
		\p_- \, &= \, \p_0 \, - \, \frac{1}{2\sqrt{2}} \log \left( \frac{1-i \sqrt{2}\g}{1+i \sqrt{2}\g} \right) \, , \qquad (\textrm{for} \ \ t<0) 
	\end{align}
Therefore, the dual CFT can be interpreted as an interface conformal field theory (ICFT) with the interface being localized in the time direction at $t=0$ and extended to all spatial directions.
(see the left panel of figure.~\ref{fig:icft} for a schematic picture of the dual ICFT.)
This interface can also be interpreted as a global quantum quench \cite{Calabrese:2006rx, Calabrese:2007rg} as we discuss below.

In order to investigate this holographic ICFT, let us now compute one-point functions.
First, we consider the dual ICFT operator of the bulk dilaton field $\p$. In the asymptotic limit $\r \to \infty$, the dilaton solution behaves as
	\begin{align}
		\p(\r) \, = \, \p_+ \, - \, \frac{2i\g}{\sqrt{1+2\g^2}} \, e^{-2\r} \, + \, \cdots \, ,
	\label{eq:phi_asympt}
	\end{align}
for $t>0$.
In the asymptotic limit, we have 
	\begin{align}
		 e^{-\r} \, \sim \, \frac{(1+2\g^2)^{\frac{1}{4}}}{2} \ \frac{z}{t} \, ,
	\label{eq:asympto-relation}
    \end{align}
where the factor $(1+2\g^2)^{\frac{1}{4}}$ will be explained later.
Therefore, the one-point function of the dual ICFT operator is 
	\begin{align}
		\big\la \mathcal{O}(t, x) \big\ra \, = \, - \, \frac{i\g}{2} \, \frac{1}{t^2} \, , \qquad (\textrm{for} \ \ t>0)
    \label{eq:<O>1}
	\end{align}
The $t<0$ result can be obtained by complex conjugation as
	\begin{align}
		\big\la \mathcal{O}(t, x) \big\ra \, = \,  \frac{i\g}{2} \, \frac{1}{t^2} \, , \qquad (\textrm{for} \ \ t<0)
    \label{eq:<O>2}
    \end{align}

Next, we also study the one-point function of the holographic ICFT stress-energy tensor.
Once we introduce the Fefferman–Graham expansion
	\begin{align}
		ds^2 \, = \, \frac{dz^2}{z^2} \, + \, \frac{dx^\m dx^\n}{z^2} \Big[ g_{(0)\m\n} + z^2 g_{(2)\m\n} + \mathcal{O}(z^4) \Big] \, ,
	\end{align}
this is computed from the bulk metric by
	\begin{align}
		\big\la T_{\m\n} \big\ra \, = \, \frac{1}{8\pi G_N} \Big[ \, g_{(2)\m\n} - g_{(0)\m\n} \, g_{(0)}^{\s\r} \, g_{(2)\s\r} \, \Big] \, .
	\end{align}
We note that the coordinates of the dual ICFT are given by $(t,x)$ as in the Poincare coordinates (\ref{eq:poincare}).
Therefore, to compute the holographic stress-energy tensor, it is useful to move back to these coordinates.
One naively thinks that this is archived by the inverse coordinate transformation of (\ref{eq:coord-transf})
which is explicitly written as $\r = \textrm{arccosh}(t/z)$ and $\h = \sqrt{t^2 - z^2 \, }$ for $t>z$.
This definition of $(t,z)$ transforms the background geometry (\ref{eq:Janus-metric}) into
	\begin{align}
		ds_3^2 \, &= \, - \, \frac{(f-1)t^2+z^2}{(t^2 - z^2)^2} \, dt^2 \, + \, \frac{2t((1+f) z^2 - t^2)}{z(t^2 - z^2)^2} \, dtdz \nn\\
        &\qquad + \, \frac{t^4 - t^2 z^2 - f z^4}{z^2(t^2 - z^2)^2} \, dz^2 \, + \, \frac{f}{t^2 - z^2} \, dx^2 \, ,
	\label{eq:ds_3}
	\end{align}
where $f = f\big( \r = \textrm{arccosh} (t/z) \big)$.
However, this is not the Gaussian normal coordinates like the Fefferman-Graham coordinates, as $g_{tz}=g_{zt}\ne 0$.
In order to archive the Gaussian normal coordinates at least asymptotically, we define the $(t,z)$ coordinates by
	\begin{align}
		\r \, = \, \textrm{arccosh} \left( \frac{t}{z} \right) \, - \, \frac{1}{4} \log(1+2\g^2) \, , \qquad \h \, = \, \sqrt{t^2 - z^2 \, } \, .
	\label{eq:inv-coord-transf}
	\end{align}
The second term in $\r$ explains the factor $(1+2\g^2)^{\frac{1}{4}}$ in (\ref{eq:asympto-relation}).
Since this is simply a shift of the $\r$ coordinate compared to the previous definition of $(t,z)$, the metric is still given by the form of (\ref{eq:ds_3}), but now $f$ is defined by
	\begin{align}
		f \, &= \, f\Big( \r = \textrm{arccosh} (t/z) \, - \, \tfrac{1}{4} \log(1+2\g^2) \Big) \nn\\[4pt]
        \, &= \, \frac{2(t^2 - z^2)+\g^2\left(2t^2 - z^2 - 2t^2 \sqrt{1-\tfrac{z^2}{t^2}}\right)}{2z^2} \, . 
	\end{align}
With this definition, we have $g_{tz}=g_{zt}= \mathcal{O}(z^3)$.
This leads to
	\begin{align}
		g_{tt} \, &= \, \frac{(1-f)t^2 - z^2}{(t^2 - z^2)^2} \, = \, - \frac{1}{z^2} \, - \, \frac{\g^2 z^2}{8t^4} \, + \, \mathcal{O}(z^4) \, , \\
		g_{xx} \, &= \, \frac{f}{t^2 - z^2} \, = \, \frac{1}{z^2} \, + \, \frac{\g^2 z^2}{8t^4} \, + \, \mathcal{O}(z^4) \, ,
	\end{align}
and $g_{tx}=g_{xt}=0$.
Therefore, the holographic CFT stress-energy tensor is 
	\begin{align}
		\big\la T_{\m\n} \big\ra \, = \, 0 \, , \qquad (\textrm{for}\ \ t \, \ne \, 0)
    \label{eq:Ttt}
	\end{align}
This result shows that energy and momentum are still conserved in each region with $t>0$ and $t<0$.
However, there is a discontinuity at $t=0$.

As in the usual space-like Janus case discussed in \cite{Clark:2004sb,Bak:2007jm}, the dual ICFT can be understood as a conformal perturbation theory 
	\begin{align}
		I_\l \, = \, I_0 \, + \, \g \int d^2x \Big( \, \th(t) \, \p_+^{(1)} + \, \th(-t) \, \p_-^{(1)} \, \Big) \mathcal{O}(t,x) \, , 
	\end{align}
where $I_0$ is the holographic CFT of empty AdS$_3$ and $\p_\pm^{(1)}$ are the linear order coefficient of $\phi_\pm$ (\ref{eq:phi_+}) for small $\g$.
In this theory, the one-point function of $\mathcal{O}$ is computed as
	\begin{align}
		\big\la \mathcal{O}(t, x) \big\ra_\g \,
        &= \, i\g \int d^2x' \Big( \, \th(t') \, \p_+^{(1)} + \, \th(-t') \, \p_-^{(1)} \, \Big) \big\la \mathcal{O}(t,x) \mathcal{O}(t', x') \big\ra_0 \, + \, \mathcal{O}(\g^2) \nn\\
		&= \, \frac{a\pi\g}{2} \bigg( \frac{\th(t) \, \p_+^{(1)} + \, \th(-t) \, \p_-^{(1)}}{t^2} \bigg) \, + \, \mathcal{O}(\g^2) \, ,
	\end{align}
where we used the two-point function in the non-perturbed theory
  	\begin{align}
		\big\la \mathcal{O}(t,x) \mathcal{O}(t', x') \big\ra_0 \, = \, \frac{a}{\big((x-x')^2 -(t-t')^2\big)^2} \, ,
	\end{align}
with some numerical coefficient $a\ne 0$.
This agrees with the bulk computation up to the overall numerical coefficient.
The next-order contribution is given by the three-point function
	\begin{align}
		\big\la \mathcal{O}(t_1, x_1) \mathcal{O}(t_2,x_2) \mathcal{O}(t_3,x_3) \big\ra_0
        \, = \, \frac{C_{\mathcal{O}\mathcal{O}\mathcal{O}}}{(x_{12}^2-t_{12}^2)(x_{23}^2-t_{23}^2)(x_{31}^2-t_{31}^2)} \, ,
	\end{align}
where $t_{ij}=t_i-t_j$ and $x_{ij}=x_i-x_j$, as
	\begin{align}
		\mathcal{O}(\g^2) \,
        &= \, -\frac{\g^2}{2} \int d^2x' \int d^2x'' \Big( \, \th(t') \, \p_+^{(1)} + \, \th(-t') \, \p_-^{(1)} \, \Big)\Big( \, \th(t'') \, \p_+^{(1)} + \, \th(-t'') \, \p_-^{(1)} \, \Big) \nn\\
        &\hspace{100pt} \times \big\la \mathcal{O}(t, x) \mathcal{O}(t',x') \mathcal{O}(t'',x'') \big\ra_0 \nn\\
        &= \, \frac{\pi^2 C_{\mathcal{O}\mathcal{O}\mathcal{O}} (\p_+^{(1)})^2}{2} \, \frac{\g^2}{t^2} \, .
	\end{align}
The detail of this computation is presented in Appendix~\ref{app:integrals}.
On the other hand, the bulk computation (\ref{eq:<O>1}), (\ref{eq:<O>2}) tells us that
there is no $\mathcal{O}(\g^2)$ contribution to the one-point function $\la \mathcal{O} \ra_\g$.
This implies that $C_{\mathcal{O}\mathcal{O}\mathcal{O}}=0$ is one of the required CFT data for the holographic dual of the time-like Janus solution.
In fact, $C_{\mathcal{O}\mathcal{O}\mathcal{O}}=0$ is also the case for the five dimensional space-like Janus solution \cite{Clark:2004sb}.

Let us also study the one-point function of the stress-energy tensor in this conformal perturbation theory.
At the first order in the deformation parameter, it is given by
	\begin{align}
		\big\la T_{\m\n}(t, x) \big\ra_\g \,
        &= \, i\g \int d^2x' \Big( \, \th(t') \, \p_+^{(1)} + \, \th(-t') \, \p_-^{(1)} \, \Big) \big\la T_{\m\n}(t, x) \mathcal{O}(t',x') \big\ra_0 \, + \, \mathcal{O}(\g^2) \nn\\
		&= \, \mathcal{O}(\g^2) \, ,
	\end{align}
where we used the vanishing two-point function $\big\la T_{\m\n}(t, x) \mathcal{O}(t',x') \big\ra_0=0$ in the non-perturbed theory \cite{Osborn:1993cr}.
The above result is consistent with the bulk computation in (\ref{eq:Ttt}).
We also look at the second order correction which is given by the three-point function as
	\begin{align}
		\big\la T_{\m\n}(t, x) \big\ra_\g \,
        &= \, -\frac{\g^2}{2} \int d^2x' \int d^2x'' \Big( \, \th(t') \, \p_+^{(1)} + \, \th(-t') \, \p_-^{(1)} \, \Big)\Big( \, \th(t'') \, \p_+^{(1)} + \, \th(-t'') \, \p_-^{(1)} \, \Big) \nn\\
        &\hspace{100pt} \times \big\la T_{\m\n}(t, x) \mathcal{O}(t',x') \mathcal{O}(t'',x'') \big\ra_0 \, + \, \mathcal{O}(\g^3) \, .
	\end{align}
The three-point function is explicitly given by \cite{Osborn:1993cr}
	\begin{align}
		\big\la T_{\m\n}(t_1, x_1) \mathcal{O}(t_2,x_2) \mathcal{O}(t_3,x_3) \big\ra_0
        \, = \, \frac{C_{T\mathcal{O}\mathcal{O}} \, h_{\m\n}(X)}{(x_{12}^2-t_{12}^2)(x_{23}^2-t_{23}^2)(x_{31}^2-t_{31}^2)} \, ,
	\end{align}
where
	\begin{align}
		h_{\m\n}(X) \, = \, \frac{X_\m X_\n}{X^2} \, - \, \frac{1}{2} \, \h_{\m\n} \, , \qquad X^\m \, = \, \frac{x_{21}^\m}{x_{21}^2} \, - \, \frac{x_{31}^\m}{x_{31}^2} \, .
	\end{align}
Without performing the integrals, we can see several facts. 
First, the trace must vanish $\big\la T^\m{}_\m(t, x) \big\ra_\g=0$, since $h_{\m\n}$ is traceless.
This is consistent with the bulk computation.
Second, the off-diagonal component vanishes $\big\la T_{tx}(t, x) \big\ra_\g=\big\la T_{xt}(t, x) \big\ra_\g=0$, since its integrand is an odd function of $x'$ or $x''$.
This is also consistent with the bulk computation.
Finally, this one-point function is symmetric under time reversal $\big\la T_{\m\n}(-t, x) \big\ra_\l = \big\la T_{\m\n}(t, x) \big\ra_\l$, since the source term is anti-symmetric under time reversal as $\p_-^{(1)} = -\p_+^{(1)}$.
To find the explicit results for the diagonal components $\big\la T_{tt}(t, x) \big\ra_\g$ and $\big\la T_{xx}(t, x) \big\ra_\g$, we need to perform the integrals.
The vanishing holographic stress tensor is consistent with conservation and scale invariance of the proposed time-interface configuration.
However, the separated-point calculation presented in Appendix~\ref{app:integrals} does not yet provide an explicit,
Ward-identity-preserving verification at second order in conformal perturbation theory.
It is important to resolve this question, but we will not investigate this point further in the present paper and leave it to future work.

\section{Holographic Entanglement Entropy}
\label{sec:entropy}

\begin{figure}[t!]
	\begin{center}
		\raisebox{16pt}{\scalebox{0.51}{\includegraphics{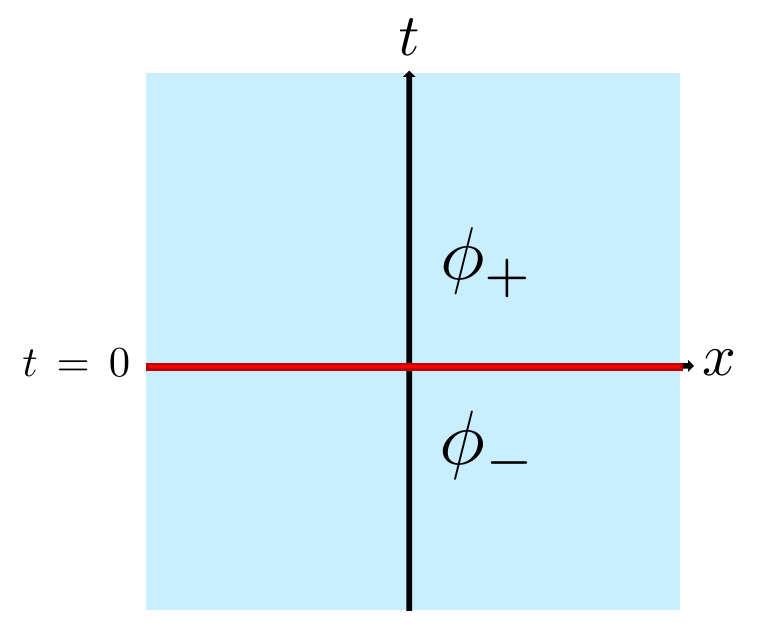}}} \qquad \scalebox{0.46}{\includegraphics{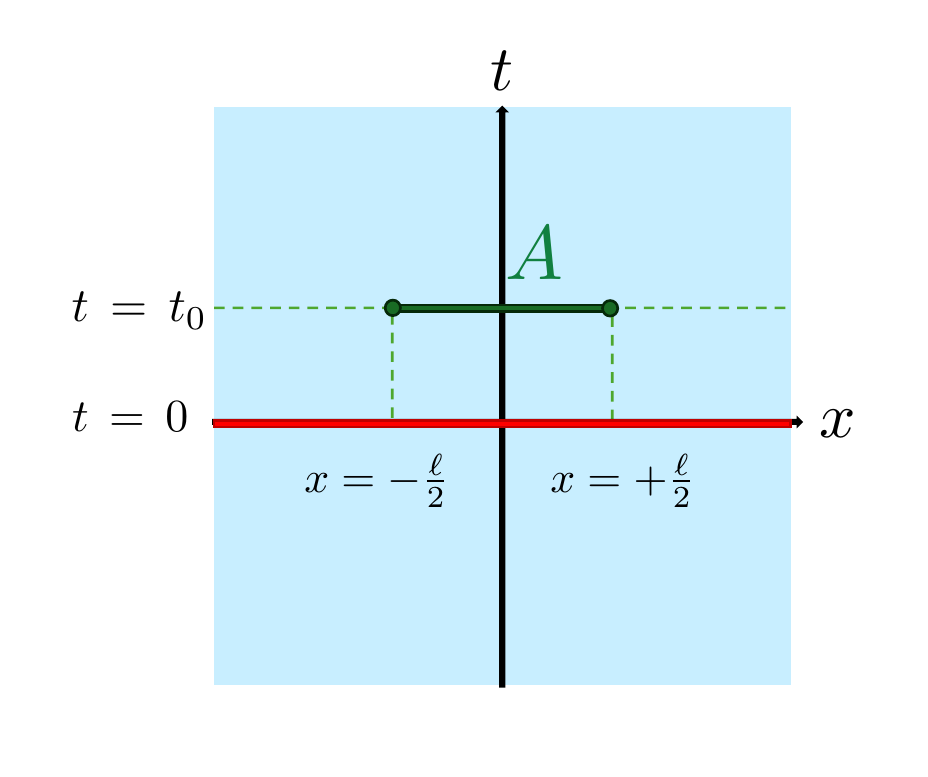}}
	\caption{(Left) a schematic picture of the dual ICFT. The interface is located at $t=0$ and extended to all spacial direction, which is depicted in red.
	(Right) a subregion of the holographic entanglement entropy we study at $t=t_0$, which is depicted in green.}
	\label{fig:icft}
	\end{center}
\end{figure}

In order to investigate further the dual ICFT, let us also study the holographic entanglement entropy now.
We consider a subregion $A$ at a spatial surface $t=t_0$ ($=$ constant) in the dual ICFT.
(In the following computation, we consider the $t_0>0$ case and will comment on the $t_0<0$ case at the end.)
Since we have translational symmetry in the $x$ direction, without loss of generality, we choose the subregion in $-\ell/2 \le x \le \ell/2$.
(See the right panel in figure~\ref{fig:icft} for the subregion $A$ in the dual ICFT.)
Since the coordinates of the dual ICFT are given by $(t,x)$, in order to compute the holographic entanglement entropy,
we need to move back to these coordinates by the inverse coordinate transformation (\ref{eq:inv-coord-transf}).
The resulting metric is given by (\ref{eq:ds_3}), and now we can clearly see that the background geometry is time-dependent.
Therefore, we need to use the HRT prescription \cite{Hubeny:2007xt} to compute the holographic entanglement entropy, instead of the static RT prescription \cite{Ryu:2006bv, Ryu:2006ef}.

The most general prescription of HRT is to find an extremal surface.
It is difficult to find an exact expression for the extremal surface, so in this section we consider small $\g$ perturbation.
In this perturbation, we expand the metric as
	\begin{align}
		g_{\m\n} \, = \, \bar{g}_{\m\n} \, + \, \g^2 \, h_{\m\n} \, + \, \mathcal{O}(\g^4) \, ,
	\label{eq:metric-expansion}
    \end{align}
where
	\begin{align}
		\bar{g}_{\m\n} dx^\m dx^\n \, &= \, \frac{-dt^2 + dx^2 +dz^2}{z^2} \, , \\[2pt]
        h_{\m\n} dx^\m dx^\n \, &= \, f_2(t,z) \left[ \frac{dx^2}{t^2 - z^2} \, - \, \frac{(t dt - z dz)^2}{(t^2-z^2)^2} \right] \, , \\[2pt]
        f_2(t,z) \, &= \, \frac{(t- \sqrt{t^2 -z^2})^2}{2z^2} \, .
	\end{align}
With this metric, the holographic entanglement entropy is computed from the geodesic distance
	\begin{align}
		L \, = \, \int d\s \sqrt{g_{\m\n}(X) \dot{X}^\m \dot{X}^\n} \, ,
	\end{align}
where $X^\m(\s)$ denotes a spacelike geodesic.
With the metric expansion (\ref{eq:metric-expansion}), we also have systematic expansion for the geodesic as
	\begin{align}
		X^\m \, &= \, \bar{X}^\m \, + \, \g^2 \, \x^\m \, + \, \mathcal{O}(\g^4) \, , \\
		L \, &= \, L_0 \, + \, \g^2 \, L_2 \, + \, \mathcal{O}(\g^4) \, .
	\end{align}
The zero-th order contribution is well known as \cite{Ryu:2006bv, Ryu:2006ef}
	\begin{align}
		L_0 \, = \, 2 \log \frac{\ell}{a} \, ,
	\end{align}
where $a$ is a UV cutoff.
In Appendix~\ref{app:extremal}, we show that the second order contribution is given by
	\begin{align}
		L_2 \, = \, \frac{1}{2} \int d\s \frac{h_{\m\n}(\bar{X}) \dot{\bar{X}}^\m \dot{\bar{X}}^\n}{\sqrt{\bar{g}_{\r\s}(\bar{X}) \dot{\bar{X}}^\r \dot{\bar{X}}^\s}} \, .
	\end{align}
Therefore, we do not need the modification of the geodesic denoted by $\x^\m$, but only need the modification of the metric by $h_{\m\n}$.
This is computed in the large $t_0$ expansion in Appendix~\ref{app:extremal} and the result is given by
	\begin{align}
		L_2 \, = \, \frac{\ell^4}{240 \, t_0^4} \, + \, \mathcal{O}\left( \frac{\ell^6}{t_0^6} \right) \, .
	\end{align}
Hence, the holographic entanglement entropy is now given by
	\begin{align}
		S_A(t_0) \, = \, \frac{L}{4G_N} \, = \, \frac{c}{3} \left( \log \frac{\ell}{a} \, + \, \frac{\g^2\ell^4}{480\, t_0^4}
        \, + \, \mathcal{O}\Big(\g^2 \ell^6 t_0^{-6}\ \, \textrm{or}\ \, \g^4 \Big) \right) \, , 
	\label{eq:S_A(t0)}
	\end{align}
where we used the Brown-Henneaux dictionary \cite{Brown:1986nw} for the central charge of the dual ICFT
	\begin{align}
		c \, = \, \frac{3}{2G_N} \, .
	\end{align}
We have been considering the $t_0>0$ case, but the metric is time reversal symmetric,
so the final result (\ref{eq:S_A(t0)}) must be valid even for the $t_0<0$ case, within the small $\g$ approximation.

\section{Stability}
\label{sec:stability}
Even though our time-like Janus solution breaks the null energy condition, 
we can show that the time-like Janus solution is stable under a probe scalar perturbation.
Therefore, in this section, we discuss the stability of the background (\ref{eq:Janus-metric}) against a probe scalar perturbation following the discussion of \cite{Bak:2003jk}.
Even though the dilaton is massless, let us start from a general massive scalar perturbation 
	\begin{align}
		0 \, &= \, - \frac{1}{\sqrt{-g}} \Big( \pa_\m \sqrt{-g} g^{\m\n} \pa_\n \p \Big) \, + \, m^2 \p \, .
	\end{align}
By assuming a separable form of solution 
	\begin{align}
		\p \, = \, e^{ikx} F(\r) G(\h) \, , 
	\end{align}
The equation can be decomposed into 
	\begin{gather}
		\Big( f^{-1} \pa_\r f \pa_\r \, - \, m^2 \Big) F(\r) \, = \, - \frac{M^2}{f} F(\r) \, , \label{eq:F-eq} \\
		\left( \pa_\h^2 \, + \, k^2 \, + \, \frac{M^2}{\h^2} \right) G(\h) \, = \, 0 \, , \label{eq:G-eq}
	\end{gather}
where $M^2$ is the separation constant.
The equation for $G(\h)$ is solved as
	\begin{align}
		G(\h) \, = \, \sqrt{\h} \, Z_{\n}(k \h) \, , \qquad \left(\n \, = \, \sqrt{\tfrac{1}{4} - M^2} \right) 
	\end{align}
where $Z_\n$ can be any Bessel function.

The energy of the perturbation is given by
	\begin{align}
		E \, = \, \frac{1}{2} \int d^2x \sqrt{-g} \Big[ - g^{\h\h} (\pa_\h \p)^2 + g^{xx} (\pa_x \p)^2 + g^{\r\r} (\pa_\r \p)^2 + m^2 \p^2 \Big] \, .
	\end{align}
Using the equation for $F(\r)$ (\ref{eq:F-eq}), this can be rewritten as
	\begin{align}
		E \, = \, \frac{1}{2} \int d^2x \left[ (\pa_\h \p)^2 + (\pa_x \p)^2 + \frac{M^2}{\h^2} \, \p^2 \right] \, .
    \label{eq:E}
	\end{align}
Now, we consider the $\r$-integral in the last term of (\ref{eq:E}):
	\begin{align}
		\int_{-\inf}^\inf d\r \, M^2 \p^2 \ \propto \ I_\r \, := \, \int_{-\inf}^\inf d\r \, f \Big[ (\pa_\r F)^2 + m^2 F^2 \Big] \, .
	\end{align}
If we consider the $\r \to + \inf$ limit, since $f \sim e^{2\r}$ in this limit, assuming $F \sim e^{\g \r}$, the equation of $F(\r)$ in (\ref{eq:F-eq}) gives $\g(\g+2)=m^2$. Therefore, the solutions are 
	\begin{align}
		\g_{\pm} \, = \, - 1 \pm \sqrt{1+m^2} \, .
	\end{align}
Only $\g_-$ gives a finite result for the $\r$-integral when the mass is above the BF bound \cite{Breitenlohner:1982bm} for AdS$_3$ (i.e. $m^2 > -1$).
If we define $v(\r):=\sqrt{f(\r)} F(\r)$, then we have 
	\begin{align}
		I_\r \, = \, \int_{-\inf}^\inf d\r \, \left[ |v'|^2 \, + \, \left( 1 + m^2 + \frac{\g^2}{2f^2} |v|^2 \right) \right]
        \, - \, \left[ \frac{f'}{2f} \, |v|^2 \right]_{-\inf}^{+\inf} \, .
	\end{align}
For the choice $\g=\g_-$, the boundary term vanishes when $m^2>-1$, and the remaining integrand is nonnegative.
Therefore, for a real probe scalar with $m^2 > -1$, we can establish a positivity of the radial operator
	\begin{align}
		I_\r \, \ge \, 0 \, ,
	\end{align}
under the boundary condition which picks up $\g_-$, 
However, we note that this discussion is merely a probe scalar perturbation analysis and this does not imply that the solution is stable under the full dilaton perturbation.

\section{Finite Temperature}
\label{sec:finite}
In this section, let us construct a time-dependent black hole, analogous to the one discussed in \cite{Bak:2007jm}.
From the metric ansatz (\ref{eq:ansatz1}), by introducing a new coordinate $\m$ by
	\begin{align}
		\r \, = \, \int_0^\m ds \sqrt{f(s)} \, ,
	\label{eq:mu}
	\end{align}
the metric becomes
	\begin{align}
		ds^2 \, = \, f(\m) \sinh^2(\m_*-\m) \, ds_{\textrm{AdS}_3}^2 \, , 
	\end{align}		
where $\m_*$ is a constant defined in (\ref{eq:mu_star}) and 
	\begin{align}		
		ds_{\textrm{AdS}_3}^2 \, = \, \frac{d\m^2 + ds_{\textrm{dS}_2}^2}{\sinh^2 (\m_*-\m)} \, .
	\label{eq:g_AdS3}
	\end{align}
As discussed in \cite{Bak:2007jm}, we can replace the AdS$_3$ metric $ds_{\textrm{AdS}_3}^2$ by the BTZ black hole metric 
	\begin{align}
		ds_{\textrm{BTZ}}^2 \, = \, - (r^2 - r_0^2) dt^2 \, + \, \frac{dr^2}{r^2 - r_0^2} \, + \, r^2 d\th^2 \, ,
	\label{eq:g_BTZ}
	\end{align}
where $0\le\th<2\pi$ and the inverse temperature of the BTZ black hole is defined by $\b=2\pi/r_0$.
Therefore, our time-dependent black hole solution is given by
	\begin{align}
		ds^2 \, = \, f(\m) \sinh^2(\m_*-\m) \, ds_{\textrm{BTZ}}^2 \, ,
	\end{align}
with $-\m_* < \m < \m_*$.
The dilaton solution is given by (\ref{eq:phi-sol}) but now $\r$ is regarded a function of $t$ and $r$ as $\r=\r(t,r)$.
This relation can be fixed as follows.
AdS$_3$ spacetime can be embedded in four-dimensional hypersurface satisfying 
	\begin{align}
		-X_0^2 + X_1^2 + X_2^2 - X_3^2 \, = \, -1 \, .
	\end{align}
The AdS$_3$ metric (\ref{eq:g_AdS3}) is obtained by parametrization 
	\begin{align}
		X_0 \, = \, \coth (\m_*-\m) \, &, \qquad
		X_1 \, = \, \frac{x}{\h \sinh (\m_*-\m)} \, , \nn\\
		X_2 \, = \, \frac{-1+x^2-\h^2}{2\h \sinh (\m_*-\m)} \, &, \qquad
		X_3 \, = \, \frac{1+x^2-\h^2}{2\h \sinh (\m_*-\m)} \, .
	\end{align}
On the other hand, the BTZ metric (\ref{eq:g_BTZ}) is obtained by parametrization 
	\begin{align}
		X_0 \, = \, \sqrt{\frac{r^2}{r_0^2} - 1} \, \sinh(r_0 t) \, &, \qquad
		X_1 \, = \, \frac{r}{r_0} \, \sinh(r_0 \th) \, , \nn\\
		X_2 \, = \, \pm \sqrt{\frac{r^2}{r_0^2} - 1} \, \cosh(r_0 t) \, &, \qquad
		X_3 \, = \, \pm \frac{r}{r_0} \, \cosh(r_0 \th) \, .
	\end{align}
Therefore, equating $X_0$, we find 
	\begin{align}
		\coth (\m_*-\m) \, = \, \sqrt{\frac{r^2}{r_0^2} - 1} \, \sinh(r_0 t) \, .
	\label{eq:mu-t-relation}
	\end{align}
Then, by (\ref{eq:mu}), $\r$ is a function of $t$ and $r$ as $\r=\r(t,r)$.

\begin{figure}[t!]
	\begin{center}
		\scalebox{0.8}{\includegraphics{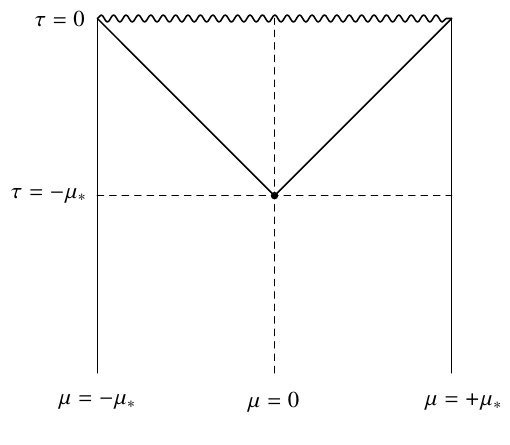}}
	\caption{A Penrose diagram of (\ref{eq:junus-bh}) for $-\inf<\t<0$ region. The vertical lines are the timelike asymptotic boundaries $\mu=\pm\mu_*$, the wavy edge is the quotient singularity at $\tau=0$, and the solid null lines are the future event horizon relative to both boundaries.}
	\label{fig:Penrose}
	\end{center}
\end{figure}

In order to see the global structure of this solution, we introduce the Kruskal coordinates as
	\begin{align}
		V \, = \, e^{r_0(t+r_*)} \, , \qquad U \, = \, -e^{-r_0(t-r_*)} \, ,
	\end{align}
where 
	\begin{align}
		r_* \, = \, \frac{1}{2r_0} \ln \left( \frac{r-r_0}{r+r_0} \right) \, ,
	\end{align}
which leads to 
	\begin{align}
		\sinh^2 (\m_*-\m) \, = \, \frac{-(1+UV)^2}{(1-U^2)(1-V^2)} \, .
	\end{align}
Then, the metric is written as
	\begin{align}
		ds^2 \, = \, \frac{f(\m)}{(1-U^2)(1-V^2)} \Big( 4 dUdV - r_0^2 (1-UV)^2 \, d\th^2 \Big) \, .
	\end{align}
The fully extended geometry may be obtained by introducing a parametrization
	\begin{align}
		V \, = \, \coth w_1 \, , \qquad U \, = \, \tanh w_2 \, ,
	\end{align}
which leads to 
	\begin{align}
		ds^2 \, &= \, f(\m) \Big( 4 dw_1dw_2 + r_0^2 \sinh^2(w_1-w_2) \, d\th^2 \Big) \nn\\
		&= \, f(\m) \Big( - d\t^2 + d\m^2 + r_0^2 \sinh^2\t \, d\th^2 \Big) \, ,
	\label{eq:junus-bh}
	\end{align}
where we introduced 
	\begin{align}
		\m \, = \, \m_* - (w_1 + w_2) \, , \qquad \t \, = \, w_2 - w_1 \, .
	\end{align}
We note that the angular orbit in $\th$ direction collapse at $\t=0$, while the scalar curvature is finite at $\t=0$. Hence, $\t=0$ surface represents a quotient singularity of the black hole solution.
The Penrose diagram of this black hole solution in the $-\inf<\t<0$ region is depicted in Figure~\ref{fig:Penrose}.

\begin{figure}[t!]
	\begin{center}
		\scalebox{0.8}{\includegraphics{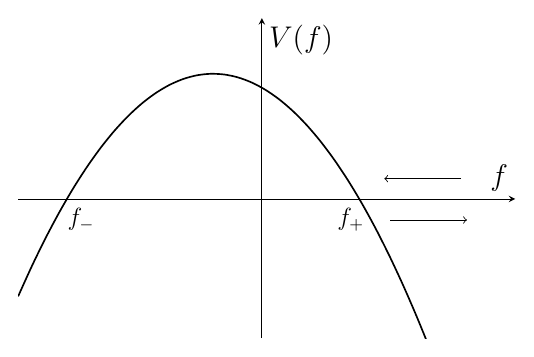}}
	\caption{A sketch of the potential $V(f)$.}
	\label{fig:V(f)}
	\end{center}
\end{figure}

Let us now study the asymptotics of $f(\m)$.
From (\ref{eq:fp-eq}) with (\ref{eq:f(rho)}), we have
	\begin{align}
		\left(\frac{df}{d\r}\right)^2 \, = \, 4\Big( f^2 + f - \tfrac{\g^2}{2} \Big) \, .
	\end{align}
As in the space-like Janus solution \cite{Bak:2003jk}, this equation has a form of the Schr\"{o}dinger equation with potential
	\begin{align}
		V(f) \, = \, - 4\Big( f^2 + f - \tfrac{\g^2}{2} \Big) \, = \, -4 (f - f_+)(f - f_-) \, ,
	\end{align}
where 
	\begin{align}
		f_{\pm} \, = \, \frac{1}{2} \Big( -1 \pm \sqrt{1+2\g^2} \Big) \, .
	\end{align}
We consider scattering from $f=\infty$ as depicted in figure~\ref{fig:V(f)}.
For $\g^2>0$, we have $f_+>0$, so that we do not get a naked singularity.
Therefore, integration of the equation leads to 
	\begin{align}
		2(\r - \r_*) \, = \, \int_{f_+}^f \frac{dx}{\sqrt{x^2 + x - \frac{\g^2}{2}}} \, ,
	\end{align}
with some integration constant $\r_*$.
For the new coordinate $\m$ defined by (\ref{eq:mu}), it is also useful to change the variables by $y=f^{-1/2}$. 
This leads to 
	\begin{align}
		\frac{dy}{d\m} \, = \, -  \sqrt{1 + y^2 - \tfrac{\g^2}{2} y^4 } \, ,
	\end{align}
and 
	\begin{align}
		\m_* - \m \, = \, \int_0^{\frac{1}{\sqrt{f}}} \frac{dy}{\sqrt{1 + y^2 - \frac{\g^2}{2} y^4}} \, .
	\label{eq:mu_star}
    \end{align}
Here $\m_*$ is chosen such that $\m=0$ corresponds to the turning point $f=f_+$. This can be explicitly seen by changing the integration variable to $u = \sqrt{f_+} y$ and writing 
	\begin{align}
		\m \, = \, \frac{1}{\sqrt{f_+}} \int_{\sqrt{f_+/f}}^1 \frac{du}{\sqrt{(1-u^2)(1- \frac{f_-}{f_+} u^2)}} \, , \qquad
		\m_* \, = \, \frac{1}{\sqrt{f_+}} \int_{0}^1 \frac{du}{\sqrt{(1-u^2)(1- \frac{f_-}{f_+} u^2)}} \, .
	\end{align}
As discussed in \cite{Freedman:2003ax}, we perform this integral by expanding the integrand around small $\g$.
	\begin{align}
		\m_* - \m \, = \, \textrm{arcsinh} (y) \, + \, \frac{\g^2 y^5}{20} \, {}_2F_1\Big( \tfrac{3}{2}, \tfrac{5}{2}; \tfrac{7}{2}; - y^2\Big) \, + \, \mathcal{O}(\g^4) \, ,
	\end{align}
where $y=f^{-1/2}$. Inverting this series, we find
	\begin{align}
		y \, = \, \sinh(\m_* - \m) \, - \, \frac{\g^2 \cosh(\m_* - \m) \sinh^5(\m_* - \m)}{20} \, {}_2F_1\Big( \tfrac{3}{2}, \tfrac{5}{2}; \tfrac{7}{2}; - \sinh^2(\m_* - \m) \Big) \, + \, \mathcal{O}(\g^4) \, .
	\end{align}
This can be rewritten as an asymptotic expansion as
	\begin{align}
		f(\m) \, \approx \, \frac{1}{\sinh^2(\m_* - \m)} \Big[ 1 + \mathcal{O}\big((\m_* - \m)^4 \big) \Big] \, .
	\end{align}
From this asymptotic behavior and (\ref{eq:fsol}), we can identify
	\begin{align}
		\frac{\sqrt{1+2\g^2}}{4} \, e^{2\r} \, \approx \, \frac{1}{\sinh^2(\m_* - \m)} \, , 
	\end{align}
where $\approx$ means that this is an asymptotic relation.
Therefore, the asymptotic behavior of the dilaton (\ref{eq:phi_asympt}) can be now written as
	\begin{align}
		\p(\r) \, &= \, \p_+ \, - \, \frac{2i\g}{\sqrt{1+2\g^2}} \, e^{-2\r} \, + \, \cdots \nn\\[2pt]
		&= \, \p_+ \, - \, \frac{i\g}{2} \, (\m_* - \m)^2 \, + \, \cdots \nn\\[2pt]
		&= \, \p_+ \, - \, \frac{i\g}{2} \left( \frac{2\pi}{\b} \right)^2 \, \frac{1}{r^2\sinh^2(\frac{2\pi t}{\b})} \, + \, \cdots \, ,
	\end{align}
where for the last equality, we used the relation (\ref{eq:mu-t-relation}).
Therefore, from this asymptotic behavior, we read off the one-point function
	\begin{align}
		\big\la \mathcal{O}(t, x) \big\ra \, = \, - \, \frac{i\g}{2} \left( \frac{2\pi}{\b} \right)^2 \, \frac{1}{\sinh^2(\frac{2\pi t}{\b})} \, .
	\end{align}

\section{General Dimension}
\label{sec:general-d}

\begin{figure}[t!]
	\begin{center}
		\scalebox{0.65}{\includegraphics{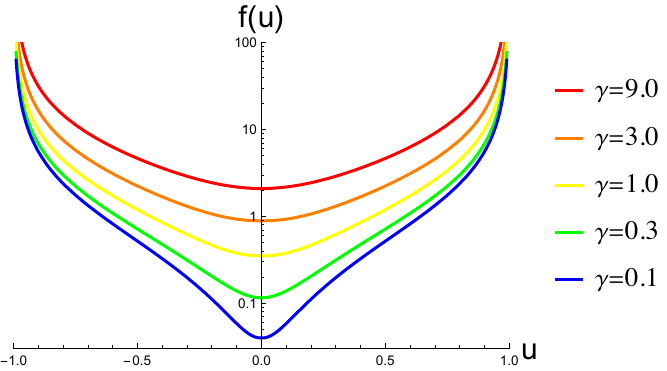} \qquad \qquad \includegraphics{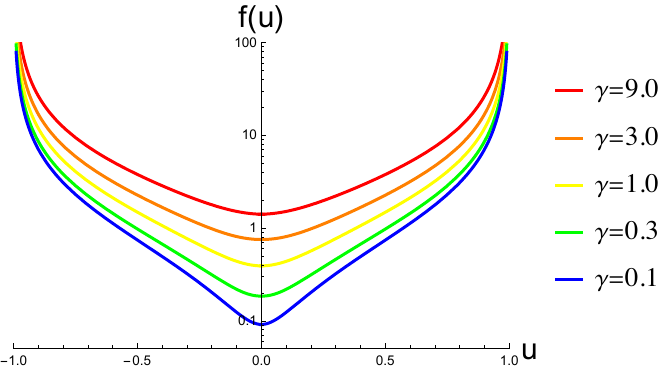}}
	\caption{Numerical solutions for $f(u)$ in $d=3$ (left) and $d=4$ (right).}
	\label{fig:f(u)}
	\end{center}
\end{figure}

In this section, we study a generalization of the zero-temperature solution discussed in section~\ref{sec:solutions} to arbitrary dimension.
We study the Einstein-dilaton theory in ($d+1$) dimensional spacetime:
	\begin{align}
		I \, = \, \frac{1}{16\pi G_N} \int d^{d+1}x \sqrt{-g} \Big[ R \, - \, 2\L \, - \, g^{\mu\nu} \pa_\mu \p \pa_\n \p \Big] \, ,
	\end{align}
where the cosmological constant is given by $\L = - \frac{d(d-1)}{2}$.
As in section~\ref{sec:solutions}, we use the ansatz 
	\begin{gather}
		ds_{d+1}^2 \, = \, d\r^2 + f(\r) ds_{\textrm{dS}_d}^2 \, , \qquad ds_{\textrm{dS}_d}^2 \, = \, \frac{-d\h^2 +d\vec{x}\,{}^2}{\h^2} \, , \\
		\p \, = \, \p(\r) \, , 
	\end{gather}
Then, the Einstein and Klein-Gordon equations are reduced to 
	\begin{gather}
		0 \, = \, 2(d-1)ff'' \, + \, \frac{(d-1)(d-4)}{2} \, f'^2 \, + \, 4 \L f^2 \, - \, 2(d-1)(d-2) f \, + \, 2 f^2 \p'^2 \, , \\[2pt]
		0 \, = \, \frac{d(d-1)}{2} \, f'^2 \, - \, 2d(d-1) f \, + \, 4 \L f^2 \, - \, 2 f^2 \p'^2 \, , \\[2pt]
		0 \, = \, \pa_\r \big(f^{\frac{d}{2}} \p' \big) \, . 
	\end{gather}
The Klein-Gordon equation leads to
	\begin{align}
		\p' \, = \, \frac{i \g}{f^{\frac{d}{2}}} \, . 
	\end{align}
With this expression, the Einstein equations are rewritten as
	\begin{gather}
		0 \, = \, ff'' \, + \, \left( \frac{d-4}{4} \right) \, f'^2 \, - \, d \, f^2 \, - \, (d-2) f \, - \, \frac{\g^2}{(d-1) f^{d-2}} \, , \\[2pt]
		0 \, = \, f'^2 \, - \, 4 f \, - \, 4 f^2 \, + \, \frac{4\g^2}{d(d-1) f^{d-2}} \, .
	\end{gather}
These equations are still difficult to analytically solve for general dimension, but it is easy to solve numerically.
We plotted the solutions for $d=3$ and $d=4$ in figure~\ref{fig:f(u)}, where we introduced 
	\begin{align}
		u \, = \, \tanh \r \, ,
	\end{align}
to make the domain compact, and defined $f(u):=f(\r)$.
%
%
If we consider asymptotic behavior, we impose the boundary condition for the metric as
	\begin{align}
		f(\r) \, \sim \, 
        \begin{cases}
        \ C_+ \, e^{2\r} \hspace{30pt} (\r \, > \, 0) \\
        \ C_- \, e^{-2\r} \qquad (\r \, < \, 0) 
        \end{cases}
	\end{align}
where $C_{\pm}$ are overall coefficients.
This leads to the dilaton behavior 
	\begin{align}
		\p(\r) \, \sim \, 
        \begin{cases}
        \ \p_+ \, - \, \dfrac{i \g}{d \, C_+^{d/2}} \, e^{-d \r} \, + \, \cdots \qquad (\r \, > \, 0) \\[10pt]
        \ \p_- \, + \, \dfrac{i \g}{d \, C_-^{d/2}} \, e^{d \r} \, + \, \cdots \hspace{30pt} (\r \, < \, 0) 
        \end{cases}
	\end{align}
As we did in section~\ref{sec:one-point}, this leads to the following one-point function
	\begin{align}
		\big\la \mathcal{O}(t, \vec{x}) \big\ra \, = \,
        \begin{cases}
        \ - \, \dfrac{i\g}{d} \, \dfrac{1}{t^d} \hspace{30pt} (t \, > \, 0) \\[10pt]
        \ + \, \dfrac{i\g}{d} \, \dfrac{1}{|t|^d} \qquad (t \, < \, 0)
        \end{cases}
	\end{align}

\section{Conclusions and Discussions}
\label{sec:conclusions}

In this paper, we have constructed a time-like Janus solution with a time-dependent dilaton field.
In order to avoid a naked singularity, we performed an analytical continuation of the Janus deformation parameter $\g$ to a pure imaginary value.
This led to a breaking of the null energy condition.
We nevertheless argued that our solution is useful as a toy model of holographic global quantum quench in the context of the AdS/ICFT correspondence.
The asymptotic value of the dilaton field approaches a constant value when $t>0$ of the ICFT and a different value when $t<0$.
Hence, the $t=0$ discontinuity can be understood as a global quench or a spacelike interface.
We computed holographic one-point functions and holographic entanglement entropy.
For the dual ICFT, we proposed a conformal perturbation theory interpretation with a global-quench-type time-dependent source.
We computed one-point functions in this conformal perturbation theory and confirmed agreement with the bulk computation.
We also discussed its stability against a probe scalar perturbation and generalization to finite temperature and to higher dimensions.

We note that our time-like Janus solution has two patches depicted in figure~\ref{fig:coordinates}.
In fact, the dilaton value coincides at the boundary of the two patches as $\p(\r=0) = \p(\hat{\r}=0) = \p_0 \pm i \pi /\sqrt{2}$, but its derivative does not.
Similarly, the metric solution is finite at the boundary of the two patches, but it does not connect smoothly.
This type of discontinuity might be understood as a shock wave propagating at the speed of light on the boundary of the two patches at $t = \pm z$.
Since a local quench \cite{Nozaki:2013wia} or a small perturbation \cite{Shenker:2013pqa} in a holographic CFT produces a shock wave in the bulk AdS spacetime,
it seems natural that our global quench also produces a shock wave in the bulk AdS spacetime.
It would be interesting to investigate this possibility further, but we leave this investigation for future work.

In this paper, we computed the holographic entanglement entropy for a later time $t_0 \gg \ell$, but not for an early time.
The CFT computation \cite{Calabrese:2005in} tells us that in the early time $t_0 < \ell/2$, the entanglement entropy grows linearly with time.
It is an important question how to derive this linear growth of the entanglement entropy from the bulk computation.
For an sufficiently early time, it seems like that the extremal surface crosses the boundary of the patch $t = \pm z$.
Therefore, it seems important to incorporate the shock wave contribution to the holographic entanglement entropy computation, if our shock wave interpretation is correct.

\section*{Acknowledgements}

We are grateful to Sumit R. Das, Dongsheng Ge, Kanato Goto, Taishi Kawamoto, Yuya Kusuki, Tatsuma Nishioka, Giuseppe Policastro, Tadashi Takayanagi and Xi Yin for useful discussion.
The work of KS is supported by JSPS KAKENHI Grant No.~23K13105.

\appendix
\section{Euclidean Solutions}
\label{app:euclidean}
In this Appendix, we study Euclidean version of the Janus solutions, and clarify the difference between the usual space-like solution and the time-like solution we are using in this paper.
We start from the Euclidean version of the Einstein-dilaton theory, and use the following ansatz 
	\begin{gather}
		ds_3^2 \, = \, d\r^2 + f(\r) ds_{H_2}^2 \, , \qquad ds_{H_2}^2 \, = \, \frac{dx^2+dy^2}{y^2} \, , \\
		\p \, = \, \p(\r) \, .
	\label{eq:metrix-ansatz-appendix}
    \end{gather}
This ansatz must contain the Euclidean version of the usual space-like Janus solution (which is used for example in \cite{Nakata:2020luh}).
With this ansatz, the Einstein and Klein-Gordon equations are reduced to 
	\begin{gather}
		2ff'' \, - \, f'^2 \, - \, 4 f^2 \, + \, 2 f^2 \p'^2 \, = \, 0 \, , \\[2pt]
		f'^2 \, + \, 4 f \, - \, 4 f^2 \, - \, 2 f^2 \p'^2 \, = \, 0 \, , \\[2pt]
		\pa_\r \big(f \p' \big) \, = \, 0 \, .
	\end{gather}
Only the second term in the second equation is changed from the Lorentzian version of the equations.
As in the Lorentzian case, the Klein-Gordon equation leads to 
	\begin{align}
		\p' \, = \, \frac{\hat{\g}}{f} \, ,
	\end{align}
with an integration constant $\hat{\g}$.
Then the remaining Einstein equations are written as
	\begin{gather}
		2ff'' \, - \, f'^2 \, - \, 4 f^2 \, + \, 2 \hat{\g}^2 \, = \, 0 \, , \\[2pt]
		f'^2 \, + \, 4 f \, - \, 4 f^2 \, - \, 2 \hat{\g}^2 \, = \, 0 \, .
	\end{gather}
These equations have two branch of solutions and the generic solutions are given by
	\begin{align}
		f(\r) \, = \, c_1 \, e^{2\r} \, + \, c_2 \, e^{-2\r} \, + \, \sqrt{4c_1 c_2 + \frac{\hat{\g}^2}{2}} \, ,
	\end{align}
where $c_1$ and $c_2$ are integration constants. The Euclidean version of the usual space-like Janus solution corresponds to the choice
	\begin{align}
		c_1 \, = \, c_2 \, = \, \frac{\sqrt{1-2\hat{\g}^2}}{4} \, .
	\end{align}
This leads to 
	\begin{align}
		f_E(\r) \, = \, \frac{1+\sqrt{1-2\hat{\g}^2} \, \cosh(2\r)}{2} \, .
	\end{align}
Then, an analytical continuation $x=i t$ leads to the standard space-like Janus solution.

On the other hand, the time-like Janus solution corresponds to a double continuation
	\begin{align}
		y \, = \, i \h \, , \qquad \r \, = \, \r_t \, + \, \frac{i\pi}{2} \, .
	\end{align}
This changes the metric ansatz (\ref{eq:metrix-ansatz-appendix}) to
	\begin{align}
		ds_3^2 \, = \, d\r_t^2 + f_T(\r_t) \left( \frac{dx^2-d\h^2}{\h^2} \right) \, ,
	\end{align}
where 
	\begin{align}
		f_T(\r_t) \, = \, - f_E\left(\r_t \, + \, \frac{i\pi}{2} \right) \, = \, \frac{-1+\sqrt{1-2\hat{\g}^2} \, \cosh(2\r_t)}{2} \, .
	\end{align}

\section{Integrals in the Conformal Perturbation Theory}
\label{app:integrals}
In this appendix we summarize some details of the integrals appeared in the conformal perturbation theory discussed in section~\ref{sec:one-point}.
In order to avoid some confusion, in this appendix, we use $y$ to denote the spacial coordinate in the CFT$_2$, and use $x=x^\m=\{t, y\}$ to denote the Lorenzian two-dimensional vector.

Let us start from the $\g$ order computation for the one-point function $\big\la \mathcal{O}(t, y) \big\ra_\g$.
This computation is basically presented in \cite{Clark:2004sb}, but for later use, we summarize it here. It is explicitly given by
	\begin{align}
		\big\la \mathcal{O}(t, y) \big\ra_\g \,
        &= \, i\g \int d^2x' \Big( \, \th(t') \, \p_+^{(1)} + \, \th(-t') \, \p_-^{(1)} \, \Big) \big\la \mathcal{O}(t,y) \mathcal{O}(t', y') \big\ra_0 \nn\\
        &= \, i a \g \int_{-\inf}^{\inf} dt' \int_{-\inf}^{\inf} dy' \, \frac{\big( \, \th(t') \, \p_+^{(1)} + \, \th(-t') \, \p_-^{(1)} \, \big)}{\big((y-y')^2 -(t-t')^2\big)^2} \, .
	\end{align}
The $y$-dependence can be absorbed in the $y'$-integral. We first perform the $y'$-integral, and then perform the $t'$-integral.
In order to make the $y'$-integral convergent, we make a Wick rotation to the Euclidean time by $t = - i \t$ during the $y'$-integral.
After the $y'$-integral, we move back to the Lorentzian time, and perform the $t'$-integral as a Lorentzian time.
Therefore, the $y'$-integral is evaluated as
	\begin{align}
		\int_{-\inf}^{\inf} \frac{dy'}{\big(y'^2 -(t-t')^2 \big)^2} \, \to \, \int_{-\inf}^{\inf} \frac{dy'}{\big(y'^2 +(\t-\t')^2 \big)^2}
        \, = \, \frac{\pi}{2|\t-\t'|^3} \, \to \, \frac{i \pi}{2|t-t'|^3} \, .
	\end{align}
There is an overall sign ambiguity in this $i \e$ prescription, but we absorb this ambiguity in the coefficient $a$.

Now, the $t'$-integral is written as
	\begin{align}
		\int_{-\inf}^{\inf} \frac{dt'}{|t-t'|^3} \, \Big( \, \th(t') \, \p_+^{(1)} + \, \th(-t') \, \p_-^{(1)} \, \Big)
        \, = \, \p_+^{(1)} \int_0^\inf \frac{dt'}{|t-t'|^3} \, + \, \p_-^{(1)} \int_{-\inf}^0 \frac{dt'}{|t-t'|^3} \, .
	\end{align}
When $t>0$, the first integral has a divergence, while when $t<0$ the second integral has a divergence.
We manage these divergences by the splitting regularization as in \cite{Clark:2004sb}.

(i) $t>0$.
	\begin{align}
		\int_0^\inf \frac{dt'}{|t-t'|^3} \, \to \, \int_{t+\d}^{\inf} \frac{dt'}{(t'-t)^3} \, + \, \int_0^{t-\d} \frac{dt'}{(t-t')^3}
        \, = \, \frac{1}{\d^2} \, - \, \frac{1}{2t^2} \, .
	\end{align}
	\begin{align}
		\int_{-\inf}^0 \frac{dt'}{|t-t'|^3} \, = \, \int_{-\inf}^0 \frac{dt'}{(t-t')^3} \, = \, \frac{1}{2t^2} \, .
	\end{align}

(ii) $t<0$.
	\begin{align}
		\int_0^\inf \frac{dt'}{|t-t'|^3} \, = \, \int_0^\inf  \frac{dt'}{(t'-t)^3} \, = \, \frac{1}{2t^2} \, .
	\end{align}    
    \begin{align}
		\int_{-\inf}^0 \frac{dt'}{|t-t'|^3} \, \to \, \int_{t+\d}^0 \frac{dt'}{(t'-t)^3} \, + \, \int_{-\inf}^{t-\d} \frac{dt'}{(t-t')^3}
        \, = \, \frac{1}{\d^2} \, - \, \frac{1}{2t^2} \, .
	\end{align}
We neglect the cutoff $\d$-dependent terms. Then the result is summarized as
	\begin{align}
		\int_{-\inf}^{\inf} \frac{dt'}{|t-t'|^3} \, \Big( \, \th(t') \, \p_+^{(1)} + \, \th(-t') \, \p_-^{(1)} \, \Big) 
        \, &= \, \frac{\th(t)}{2t^2} \, \big( -\p_+^{(1)} + \p_-^{(1)} \big) \, + \, \frac{\th(-t)}{2t^2} \, \big( \p_+^{(1)} - \p_-^{(1)} \big) \nn\\[2pt]
        &= \, - \, \frac{\big( \, \th(t) \, \p_+^{(1)} + \, \th(-t) \, \p_-^{(1)} \, \big)}{t^2}\, ,
    \end{align}
where we used the fact that $\p_-^{(1)} = -\p_+^{(1)}$.
Therefore, the one-point function is given by
	\begin{align}
		\big\la \mathcal{O}(t, y) \big\ra_\g \,
		&= \, \frac{a\pi\g}{2} \bigg( \frac{\th(t) \, \p_+^{(1)} + \, \th(-t) \, \p_-^{(1)}}{t^2} \bigg) \, + \, \mathcal{O}(\g^2) \, .
	\end{align}

Next, we proceed to the $\mathcal{O}(\g^2)$ computation for this one-point function. This is explicitly given by
	\begin{align}
		\big\la \mathcal{O}(t_1, y_1) \big\ra_\g \, &\supset \, -\frac{\g^2}{2} \int d^2x_2 d^2x_3
        \Big( \, \th(t_2) \, \p_+^{(1)} + \, \th(-t_2) \, \p_-^{(1)} \, \Big)\Big( \, \th(t_3) \, \p_+^{(1)} + \, \th(-t_3) \, \p_-^{(1)} \, \Big) \nn\\
        &\hspace{100pt} \times \big\la \mathcal{O}(t_1,y_1) \mathcal{O}(t_2,y_2) \mathcal{O}(t_3,y_3) \big\ra_0 \\[4pt]
        &= \, -\frac{C_{\mathcal{O}\mathcal{O}\mathcal{O}} \g^2}{2} \int d^2x_2 d^2x_3
        \frac{\big( \, \th(t_2) \, \p_+^{(1)} + \, \th(-t_2) \, \p_-^{(1)} \, \big)\big( \, \th(t_3) \, \p_+^{(1)} + \, \th(-t_3) \, \p_-^{(1)} \, \big)}
        {(y_{12}^2-t_{12}^2)(y_{23}^2-t_{23}^2)(y_{31}^2-t_{31}^2)} \, . \nn
	\end{align}
The $y_2$ and $y_3$ integrals are given by the following form of integral as
	\begin{align}
		\int_{-\inf}^\inf \frac{dy_c}{(y_{ac}^2-t_a^2)(y_{bc}^2-t_b^2)} \, = \, \frac{i \pi (|t_a|+|t_b|)}{|t_a||t_b|\big( (|t_a|+|t_b|)^2 - y_{ab}^2 \big)} \, ,
	\end{align}
where we used the Wick rotation as before.
Then, the remaining $t$-integrals are given by
	\begin{align}
		&\qquad \big\la \mathcal{O}(t_1, y_1) \big\ra_\g \nn\\
        \, &\supset \, -\frac{\pi^2 C_{\mathcal{O}\mathcal{O}\mathcal{O}} \g^2}{2} \int_{-\inf}^\inf dt_2 \int_{-\inf}^\inf dt_3
        \frac{\big( \, \th(t_2) \, \p_+^{(1)} + \, \th(-t_2) \, \p_-^{(1)} \, \big)\big( \, \th(t_3) \, \p_+^{(1)} + \, \th(-t_3) \, \p_-^{(1)} \, \big)}
        {|t_{12}||t_{23}||t_{31}|(|t_{12}|+|t_{23}|+|t_{31}|)} \nn\\
        &= \, -\frac{\pi^2 C_{\mathcal{O}\mathcal{O}\mathcal{O}} \g^2}{2}
        \left[ - \frac{(\p_+^{(1)})^2}{4t_1^2} + \frac{\p_+^{(1)}\p_-^{(1)}}{4t_1^2} + \frac{\p_-^{(1)}\p_+^{(1)}}{4t_1^2} - \frac{(\p_-^{(1)})^2}{4t_1^2} \right] \nn\\
        &= \, \frac{\pi^2 C_{\mathcal{O}\mathcal{O}\mathcal{O}}(\p_+^{(1)})^2}{2} \frac{\g^2}{t_1^2} \, ,
	\label{eq:t-integral}
    \end{align}
where we used the splitting regularization as before and $\p_-^{(1)} = -\p_+^{(1)}$.

Finally, we study the stress-energy tensor one-point function at $\mathcal{O}(\g^2)$ order. This is given by
	\begin{align}
		\big\la T_{\m\n}(t_1, y_1) \big\ra_\g \,
        &= \, -\frac{\g^2}{2} \int d^2x_2 \int d^2x_3 \Big( \, \th(t_2) \, \p_+^{(1)} + \, \th(-t_2) \, \p_-^{(1)} \, \Big)\Big( \, \th(t_3) \, \p_+^{(1)} + \, \th(-t_3) \, \p_-^{(1)} \, \Big) \nn\\
        &\hspace{100pt} \times \big\la T_{\m\n}(t_1,y_1) \mathcal{O}(t_2,y_2) \mathcal{O}(t_3,y_3) \big\ra_0 \, + \, \mathcal{O}(\g^3) \, .
	\end{align}
The three-point function is explicitly given by \cite{Osborn:1993cr}
	\begin{align}
		\big\la T_{\m\n}(t_1, y_1) \mathcal{O}(t_2,y_2) \mathcal{O}(t_3,y_3) \big\ra_0
        \, = \, \frac{C_{T\mathcal{O}\mathcal{O}} \, h_{\m\n}(X)}{(y_{12}^2-t_{12}^2)(y_{23}^2-t_{23}^2)(y_{31}^2-t_{31}^2)} \, ,
	\end{align}
where
	\begin{align}
		h_{\m\n}(X) \, = \, \frac{X_\m X_\n}{X^2} \, - \, \frac{1}{2} \, \h_{\m\n} \, , \qquad X^\m \, = \, \frac{x_{21}^\m}{x_{21}^2} \, - \, \frac{x_{31}^\m}{x_{31}^2} \, .
	\end{align}
We first note that the $y_1$-dependence only appears through $y_{21}$ and $y_{31}$, so by shifting the integration variables, the result is independent of $y_1$.
Furthermore, since we can express 
	\begin{align}
		X^2 \, = \, \frac{x_{23}^2}{x_{21}^2 x_{31}^2} \, , 
	\end{align}
$h_{ty}(X)=h_{yt}(X)$ is an odd function of $y_2$ or $y_3$. This leads to the vanishing off-diagonal components $\la T_{ty}\ra_\g = \la T_{yt}\ra_\g = 0$.
We also note that by construction \cite{Osborn:1993cr} the three-point function is traceless $h^{\m}{}_\m=0$.
This leads to $\la T^{\m}{}_\m \ra_\g = 0$, so for the diagonal components, we only need to compute one of them.
The integrations for $T_{yy}$ are simpler, so we compute $\la T_{yy}\ra_\g$ here.
There are two contributions for this one-point function; the first comes from the first term in $h_{yy}(X)$ and the second comes from the second term in $h_{yy}(X)$.
We denote these contributions by $\la T_{yy}\ra_\g^{(1)}$ and $\la T_{yy}\ra_\g^{(2)}$, respectively.
The integrations for $\la T_{yy}\ra_\g^{(2)}$ are equivalent to those evaluated for the $\mathcal{O}(\g^2)$ order of $\la \mathcal{O} \ra_\g$ above.
Therefore, we only need to evaluate $\la T_{yy}\ra_\g^{(1)}$ here.
This contribution is written as
	\begin{align}
		\big\la T_{yy}(t_1, y_1) \big\ra_\g^{(1)} \,
        &= \, - \frac{\g^2 C_{T\mathcal{O}\mathcal{O}}}{2} \int d^2x_2 \int d^2x_3 \Big( \, \th(t_2) \, \p_+^{(1)} + \, \th(-t_2) \, \p_-^{(1)} \, \Big) \\
        & \quad \times \Big( \, \th(t_3) \, \p_+^{(1)} + \, \th(-t_3) \, \p_-^{(1)} \, \Big)
        \frac{(x_{31}^2 y_2 - x_{21}^2 y_3)^2}{x_{12}^4 x_{23}^4 x_{31}^4} \, + \, \mathcal{O}(\g^3) \, . \nn
	\end{align}
The $y_2$ and $y_3$ integrals are evaluated by using the Wick rotation as before. This leads to 
	\begin{align}
		&\big\la T_{yy}(t_1, y_1) \big\ra_\g^{(1)} \,
        = \, - \frac{\pi^2 C_{T\mathcal{O}\mathcal{O}} \g^2}{8} \int_{-\inf}^\inf dt_2 \Big( \, \th(t_2) \, \p_+^{(1)} + \, \th(-t_2) \, \p_-^{(1)} \, \Big) \\
        &\qquad \qquad \times \int_{-\inf}^\inf \frac{dt_3}{|t_{23}|^3} \Big( \, \th(t_3) \, \p_+^{(1)} + \, \th(-t_3) \, \p_-^{(1)} \, \Big)
        \left[ \frac{1}{|t_{12}|} \, + \, \frac{1}{|t_{13}|} \, - \, \frac{4(|t_{12}|+|t_{13}|+2|t_{23}|)}{(|t_{12}|+|t_{13}|+|t_{23}|)^2} \right] \, . \nn
	\end{align}
We can simplify the integration kernel before evaluating the remaining $t$ integrals by using the following identity:
	\begin{align}
		&\qquad \frac{1}{|t_{23}|^3}
        \left[ \frac{1}{|t_{12}|} \, + \, \frac{1}{|t_{13}|} \, - \, \frac{4(|t_{12}|+|t_{13}|+2|t_{23}|)}{(|t_{12}|+|t_{13}|+|t_{23}|)^2} \right] \\
        &= \, \frac{2}{|t_{12}||t_{13}||t_{23}|(|t_{12}|+|t_{13}|+|t_{23}|)}
        \, - \, \frac{3}{|t_{23}|^4} \Big( \th(t_{12})\th(t_{31}) + \th(t_{21})\th(t_{13}) \Big) \, . \nn
	\end{align}
The $t$ integrals from the first term is already evaluated in (\ref{eq:t-integral}). Combining the result from the second term, we find
	\begin{align}
		\big\la T_{yy}(t_1, y_1) \big\ra_\g^{(1)} \, = \, 0 \, + \, \mathcal{O}(\g^3) .
	\end{align}
Combining the $\la T_{yy}\ra_\g^{(2)}$ contribution, we finally find 
	\begin{align}
		\big\la T_{yy}(t_1, y_1) \big\ra_\g \, = \, - \frac{\pi^2 C_{T\mathcal{O}\mathcal{O}} (\p_+^{(1)})^2}{4} \, \frac{\g^2}{t_1^2} \, + \, \mathcal{O}(\g^3)\, .
	\end{align}
Then the traceless condition $\la T^{\m}{}_\m \ra_\g = - \la T_{tt} \ra_\g + \la T_{yy} \ra_\g = 0$ gives 
	\begin{align}
		\big\la T_{tt}(t_1, y_1) \big\ra_\g \, = \, - \frac{\pi^2 C_{T\mathcal{O}\mathcal{O}} (\p_+^{(1)})^2}{4} \, \frac{\g^2}{t_1^2} \, + \, \mathcal{O}(\g^3) \, .
	\end{align}

\section{Extremal Surface}
\label{app:extremal}
In this appendix we discuss the co-dimenion two extremal surface in the time-dependent spacetime (\ref{eq:ds_3}).
As discussed in section~\ref{sec:entropy}, this is given by a geodesic anchored at $(t,x,z)=(t_0, \pm\ell/2, a)$, where $a$ is a UV cutoff.
Then, the geodesic distance is given by
	\begin{align}
		L \, = \, \int d\s \sqrt{g_{\m\n}(X) \dot{X}^\m \dot{X}^\n} \, ,
	\end{align}
where $X^\m(\s)$ denotes the spacelike geodesic.
We compute this geodesic distance in the small $\g$ expansion.
Therefore, we expand the metric as
	\begin{align}
		g_{\m\n} \, = \, \bar{g}_{\m\n} \, + \, \g^2 \, h_{\m\n} \, + \, \mathcal{O}(\g^4) \, ,
    \end{align}
where
	\begin{align}
		\bar{g}_{\m\n} dx^\m dx^\n \, &= \, \frac{-dt^2 + dx^2 +dz^2}{z^2} \, , \\[2pt]
        h_{\m\n} dx^\m dx^\n \, &= \, f_2(t,z) \left[ \frac{dx^2}{t^2 - z^2} \, - \, \frac{(t dt - z dz)^2}{(t^2-z^2)^2} \right] \, , \\[2pt]
        f_2(t,z) \, &= \, \frac{(t- \sqrt{t^2 -z^2})^2}{2z^2} \, .
	\end{align}
With this metric expansion, we also have systematic expansion for the geodesic as
	\begin{align}
		X^\m \, &= \, \bar{X}^\m \, + \, \g^2 \, \x^\m \, + \, \mathcal{O}(\g^4) \, , \\
		L \, &= \, L_0 \, + \, \g^2 \, L_2 \, + \, \mathcal{O}(\g^4) \, .
	\end{align}

The zero-th order geodesic is given by
	\begin{align}
		t \, =\, t_0 \, , \qquad z(x) \, = \, \sqrt{(\ell/2)^2 + a^2 - x^2} \, .
	\end{align}
Hence, the geodesic distance is
	\begin{align}
		L_0 \, &= \, \int_{-\ell/2}^{\ell/2} dx \, \frac{\sqrt{(\ell/2)^2 + a^2}}{(\ell/2)^2 + a^2 - x^2} \nn\\
        &= \, \log\left( \frac{\sqrt{(\ell/2)^2 + a^2} + \ell/2}{\sqrt{(\ell/2)^2 + a^2} - \ell/2} \right) \nn\\
        &= \, 2 \log \left( \frac{\ell}{a} \right) \, + \, \mathcal{O}\left( \frac{a^2}{\ell^2} \right) \, .
	\end{align}
where $a$ is a UV cutoff.

Next we study the second order contribution
	\begin{align}
		L_2 \, = \, \frac{dL}{d\g^2} \bigg|_{\g^2=0} \, .
	\end{align}
Since the metric is evaluated on the moving curve, its total derivative contains both its explicit $\g^2$-dependence and the change of its argument:
	\begin{align}
		\frac{d g_{\m\n}(X)}{d\g^2} \bigg|_{\g^2=0} \, = \, h_{\m\n}(\bar{X}) \, + \, \x^\r \, \pa_\r \bar{g}_{\m\n}(\bar{X}) \, .
	\end{align}
Then, defining 
	\begin{align}
		e_0 \, = \, \sqrt{\bar{g}_{\m\n}(\bar{X}) \dot{\bar{X}}^\m \dot{\bar{X}}^\n} \, ,
	\end{align}
we can evaluate the second order contribution as
	\begin{align}
		L_2 \, &= \, \int_{\G_0} \frac{d\s}{e_0} \left[ \frac{1}{2} \, h_{\m\n} \dot{\bar{X}}^\m \dot{\bar{X}}^\n \, + \, \frac{1}{2} \, \x^\r \, \pa_\r \bar{g}_{\m\n}(\bar{X}) \dot{\bar{X}}^\m \dot{\bar{X}}^\n \, + \, \bar{g}_{\m\n} \dot{\bar{X}}^\m \dot{\x}^\n \right] \nn\\
        &= \, \int_{\G_0} \frac{d\s}{e_0} \left[ \frac{1}{2} \, h_{\m\n} \dot{\bar{X}}^\m \dot{\bar{X}}^\n \, + \, \dot{\bar{X}}^\m \bar{g}_{\m\n} (\bar{\nabla} \x)^\n \right] \label{eq:L2-formal} \\
        &= \, \int_{\G_0} \frac{d\s}{e_0} \left[ \frac{1}{2} \, h_{\m\n} \dot{\bar{X}}^\m \dot{\bar{X}}^\n \, - \, \bar{g}_{\m\n} (\bar{\nabla} \dot{\bar{X}})^\m \x^\n \right] \, + \, \left[ \frac{\bar{g}_{\m\n} \dot{\bar{X}}^\m \x^\n}{e_0} \right]_{\pa \G_0} \, , \nn
	\end{align}
where $\G_0$ is the zero-th order geodesic and $\bar{\nabla}$ denote the Levi–Civita connection of $\bar{g}_{\m\n}$:
	\begin{align}
		(\bar{\nabla} \x)^\n \, = \, \frac{d \x^\n}{d\s} \, + \, \bar{\G}^\n_{\r\s} \dot{\bar{X}}^\r \x^\s \, .
	\end{align}
Now in (\ref{eq:L2-formal}) the second term vanishes because $\G_0$ is the zero-th order geodesic and the last term also vanishes due to the boundary conditions.
Hence, only the first term survives and the second order contribution is written as
	\begin{align}
		L_2 \, = \, \frac{1}{2} \int d\s \frac{h_{\m\n}(\bar{X}) \dot{\bar{X}}^\m \dot{\bar{X}}^\n}{\sqrt{\bar{g}_{\r\s}(\bar{X}) \dot{\bar{X}}^\r \dot{\bar{X}}^\s}} \, .
	\end{align}
Therefore, we do not need the modification of the geodesic denoted by $\x^\m$, but only need the modification of the metric by $h_{\m\n}$.
This integral is finite in the $a\to 0$ limit and it is computed as
	\begin{align}
		L_2 \, &= \, \frac{t_0^2 - (\ell/2)^2}{\ell} \int_{-\ell/2}^{\ell/2} dx \, \frac{z^2(x) f_2(t_0, z(x))}{(t_0^2 - z^2(x))^2} \nn\\
        &= \, \frac{1}{8\ell t_0^4} \int_{-\ell/2}^{\ell/2} dx \, \big( (\ell/2)^2 - x^2 \big)^2 \, + \, \mathcal{O}\left( \frac{\ell^6}{t_0^6} \right) \nn\\
        &= \, \frac{\ell^4}{240 t_0^4} \, + \, \mathcal{O}\left( \frac{\ell^6}{t_0^6} \right) \, .
	\end{align}

\bibliographystyle{JHEP}
\bibliography{Refs}


\end{document}